\documentstyle[11pt]{article}
\catcode`\@=11 
\def\lsim{\mathrel{\mathpalette\@versim<}}
\def\gsim{\mathrel{\mathpalette\@versim>}}
\def\@versim#1#2{\vcenter{\offinterlineskip
        \ialign{$\m@th#1\hfil##\hfil$\crcr#2\crcr\sim\crcr } }}
\catcode`\@=12 
\begin{document}
\date{}
\title{Hot One-Temperature Accretion Flows Around Black Holes}
\author{Ann A. Esin, Ramesh Narayan, Eve Ostriker \\
Harvard-Smithsonian Center for Astrophysics, Cambridge, MA 02138 \\ \\
Insu Yi \\ Institute for Advanced Study, Princeton, NJ 08540}
\maketitle
\begin{abstract}
We describe hot, optically-thin solutions for one-temperature
accretion disks around black holes.  We include cooling by synchrotron,
bremsstrahlung, and Comptonization.  Our solutions are thermally and
viscously stable, with gas temperatures on the order of $T \sim
10^9-10^{10.7}$K.  The thermal stability is a direct result of the inclusion 
of synchrotron cooling.

The new solution branch is related to the advection-dominated solution
for a two-temperature gas described by Narayan \& Yi (1995b).  It is
present only for mass accretion rates less than some critical
$\dot{M}_{crit}$ which depends on the radius $R$ and viscosity
parameter $\alpha$.  The solutions are advection-dominated for
extremely low values of $\dot{M}$.  However, for a range of
intermediate accretion rates, the new solutions are both hot ($T \sim
10^{10}$K) and cooling-dominated.  Because of this new feature,
one-temperature solutions are significantly more luminous than the 
corresponding two temperature solutions.  
 
The radial profile of the new solutions is unusual.  The inner parts of the
flow are cooling-dominated and have a disk-like geometry, while the outer 
parts are fully advection-dominated and nearly quasi-spherical.
\end{abstract}

\section{Introduction}

The best known model of an accretion flow is the standard thin
accretion disk developed by Shakura \& Sunyaev (1973), Novikov \&
Thorne (1973), and Lynden-Bell \& Pringle (1974) (see Frank, King \&
Raine (1992) for a detailed discussion).  The basic assumption of the
model is that the accreting gas is cool, compared to the local virial
temperature, so that the flow acquires a thin disk configuration.  The
disk parameters are calculated assuming an equilibrium between the
viscous energy generation inside the disk and radiative cooling from
the surface; the latter is computed under the assumption that the disk
is optically thick in the vertical direction.

The thin disk model has been widely and successfully used for modeling
various low energy systems such as accreting white dwarfs and
pre-main-sequence stars (Frank et al. 1992).  However, the model has
had less success explaining the characteristics of relativistic
objects such as accreting black holes.  There are essentially two
problems: 1) In a system containing an accreting black hole, the
accretion disk must extend inwards to the Schwarzschild radius,
$R_{Schw}$.  However, for reasonable mass accretion rates and the
standard $\alpha$ viscosity prescription, the Lightman-Eardley
instability (Lightman \& Eardley 1974) causes a breakdown of the thin
disk configuration at radii $R\sim R_{Schw}$.  Therefore, there
is some doubt that a thin accretion disk can exist at all close to a
black hole.  2) Even if a thin disk is viable, there is a problem with
the spectrum.  Since the gas in a thin disk model is optically thick,
we would expect a roughly blackbody-like spectrum at the high frequency end,
with a cutoff at several keV.
However, extensive observations of black hole systems by GINGA, GRANAT, 
and GRO (Tanaka 1989, Grebenev et al. 1993,
Maisack et al. 1993, Harmon et al. 1994, Johnson et al. 1994, Kinzer
et al. 1994, Gilfanov et al. 1995) have made it clear that the
spectra of most black hole accretors have a hard power-law components
extending to few$\times$100 keV.  This component has to be radiated by
an optically thin plasma with temperature $T \gsim 10^9$ K.  The
required temperature is at least an order of magnitude higher than that 
predicted by the standard thin disk model.

Shapiro, Lightman, \& Eardley (1976, hereafter SLE) discovered a new
class of solutions at sub-Eddington accretion rates, where the
accretion flow is optically thin and quite hot ($T_e \sim 10^8-10^9$
K).  These authors also introduced the important idea of a
two-temperature plasma, in which ions are much hotter than electrons.
Since the SLE solution has exactly the necessary characteristics
(optically thin hot gas) to explain the spectra of accreting black
holes, it has been widely studied and applied in models of X-ray binaries
and active galactic nuclei (e.g. Kusunose \& Takahara 1985, 1989; White 
\& Lightman 1989; Wandel \& Liang 1991; Melia \& Misra 1993; Luo \& Liang 
1994).  Unfortunately, the SLE solution is thermally unstable (Pringle, Rees,
\& Pacholczyk 1973, Piran 1978).  The efficiency of bremsstrahlung
cooling in the optically thin gas is proportional to the particle
density which decreases with increasing temperature.  Therefore, as
the gas is perturbed to a higher temperature, the rate of cooling per
unit mass decreases and the gas heats up even further in a runaway
process.  The thermal instability of the SLE solution makes it
unlikely that real flows can take up this configuration.

Recently, a new class of two-temperature advection-dominated solutions
has been discovered (Narayan \& Yi 1994, 1995a, 1995b; Abramowicz et
al. 1995; Chen 1995; Chen et al. 1995).  These advection-dominated
solutions are optically thin, are hotter even than the SLE solution,
and are viscously and thermally stable to large-wavelength
perturbations (Abramowicz et al. 1995, Narayan \& Yi 1995b).  Although
Kato, Abramowicz \& Chen (1995) have discovered an instability in
these flows at short wavelengths, they show that the mode amplitude
does not grow significantly and so the viability of the solutions is
not in doubt.  The advection-dominated model has been fairly
successful in explaining a number of low-luminosity systems (e.g. a
model for Sagittarius A$^*$ by Narayan, Yi, \& Mahadevan [1995], a
model for the soft X-ray transient A0620-00 by Narayan, McClintock \&
Yi [1996], and a model for the central source in NGC 4258 by Lasota et
al. [1996]) and is perhaps also a reasonable model for more luminous
systems (Narayan 1996; Narayan, Yi, \& Mahadevan 1996).

The most detailed advection-dominated models considered so far are
based on a two-temperature plasma which cools via bremsstrahlung,
synchrotron and Comptonization processes.  However, there have been
discussions in the literature questioning whether a two temperature
plasma can occur at all in astrophysical accretion flows (Phinney
1981, Rees et al. 1982).  The discussion has been further fueled by
the work of Begelman \& Chiueh (1988), who identified a particular
mechanism involving collective plasma waves, which may under certain
conditions pump energy directly into the electrons and bring the ions
and electrons into thermal equilibrium.  An important question
therefore is the following: are there stable hot solutions, analogous to
the two-temperature advection-dominated solution, if the
plasma is well-coupled and has a single temperature for the ions and
electrons?  Abramowicz et al. (1995) and Chen (1995) did consider
single-temperature plasmas in their studies of advection-dominated flows, 
but they included only
bremsstrahlung cooling.  As yet, one-temperature models with a more
detailed cooling, including synchrotron radiation and Comptonization
of synchrotron and bremsstrahlung photons, has not been considered,
even though at $T \sim 10^{10}$ K with an equipartition magnetic field
these processes completely
dominate over pure bremsstrahlung cooling.  This is the study we
present here.

In \S2 we derive the equations that describe an accreting
one-temperature plasma.  We then 
present the main results of the paper in \S3.  We find that, as in
the two-temperature case (Narayan \& Yi 1995b, hereafter NY), the
equations allow three branches of solutions: the standard thin Shakura
\& Sunyaev disk, a hot optically-thin, thermally unstable solution,
which is equivalent to the SLE branch for a two-temperature plasma,
and finally a new hotter and thermally stable disk solution.  We
describe the properties of the new solution and compare it to the
corresponding two-temperature solution branch.  Among other things, we
find that the one-temperature solution is hotter than the equivalent
two-temperature one and is {\it not} advection-dominated for a wide
range of accretion rates.  This means that for a given $\dot{M}$ the
one-temperature disk is considerably more luminous.  We conclude in
\S4 with a summary and discussion.

\section{Basic Equations}

\subsection{One-Temperature Accretion Flow}
Narayan \& Yi (1994, 1995b) derived a set of local equations
describing a vertically averaged, axisymmetric, two-temperature
advection-dominated flow.  Here we slightly modify these equations by
setting the electron and ion temperatures equal to each other.  We
also include radiation pressure, which NY neglected.

The vertically averaged pressure $p$ and density $\rho$ of the accreting gas
are given by:
\begin{equation}
\label{2.1}
p = \rho c_s^2, \ \ \ \ \rho = \frac{\dot{M}}{4 \pi R H |v|},
\end{equation}
where $c_s$ is the local isothermal sound speed, $\dot{M}$ is the mass
accretion rate, $R$ is the radius, $H$ is the vertical scale height of
the disk, and $v$ is the radial velocity.  We write the total pressure
as the sum of the gas, radiation and magnetic pressures, $p = p_g +
p_r + p_m$, and we define three quantities that describe their relative
magnitudes: $\beta = p_g/p$ determines the fraction of the total
pressure due to gas pressure; $\beta_m = (p_r + p_g)/p = 1 - p_m/p$
determines the importance of magnetic field pressure; finally $\chi$
is defined such that $\beta = \chi \beta_m$ so that $\chi$ represents
the ratio of gas pressure to the sum of gas and radiation pressure, 
$\chi = p_g/(p_g+p_r)$.
For $\chi \sim 1$ radiation pressure is unimportant, and for $\chi \ll
1$ radiation pressure dominates over gas pressure.  In our
calculations we fix the value of $\beta_m$, typically at $0.5$
corresponding to magnetic pressure equal to $50 \%$ of the total
pressure, and then solve for $\chi$ and $\beta$.

With the above definitions, the scaled relations derived by NY take
the form:
\begin{eqnarray}
\label{2.2}
v & = & -2.12\times 10^{10}\ \frac{\alpha c_1}{\sqrt{r}}\ 
{\rm cm\,s^{-1}}, \nonumber \\
c_s^2 & = & 4.5\times 10^{20}\ \frac{c_3}{r}\ {\rm cm^2\,s^{-2}}, \nonumber \\ 
\rho & = & 3.79\times 10^{-5} \frac{\dot{m}}{\alpha c_1 m \sqrt{c_3 r^3}}\  
{\rm g\,cm^{-3}}, \nonumber \\
B & = & 6.55\times 10^8 \sqrt{\frac{(1-\beta_m) \dot{m}}{\alpha c_1 m}
\sqrt{\frac{c_3}{r^5}}}\ \mbox{G}, \nonumber \\
f&=&\frac{1}{\epsilon^{\prime}} \left(\frac{5/3-\gamma}{\gamma-1}\right), 
\ \ \ \ \gamma = \frac{32-24 \beta - 3\beta^2}{24 - 21\beta}, \nonumber \\
q^+ & = & 1.84\times 10^{21}\ \frac{\epsilon^{\prime} \dot{m} \sqrt{c_3}}
{m^2 r^4}\ {\rm erg\,cm^{-3}\,s^{-1}}.
\end{eqnarray}
Here $B$ is the magnetic field strength and the quantity $f$
determines the degree to which the flow is advection-dominated. $f$
ranges from 0 to 1; the limit of $f = 1$ corresponds to the extreme
case when all the energy is stored in the gas and advected, while $f
\rightarrow 0$ corresponds to a standard cooling-dominated disk where
most of the generated energy is radiated locally.  The quantity $q^+$
is the viscous dissipation of energy per unit volume, and the
factors $\epsilon^{\prime}, c_1$, and $c_3$ are given by
\begin{eqnarray}
\label{2.3}
\epsilon^{\prime}&=&\frac{1}{2} \left(\frac{18 \alpha^2 - x^2}{2 x}-5\right),
\ \ \ \ x = \frac{9 c_3 \alpha^2}{2}, \nonumber \\
c_1&=&\frac{3}{2} c_3, \nonumber \\
c_3&=&3.12\times 10^{-13}\ \frac{T r}{\beta}.
\end{eqnarray}

In the above equations masses are scaled in units of the solar mass  
\[M = m M_{\odot};\]
accretion rates in Eddington units
\[\dot{M} = \dot{m} \dot{M}_{Edd},\ \ \ \ \dot{M}_{Edd} = \frac{L_{Edd}}
{\eta_{eff} c^2} = \frac{4 \pi G M}{\eta_{eff} \kappa_{es} c} = 1.39
\times 10^{18} m\ {\rm g\,s^{-1}},\] 
where $\kappa_{es} = 0.4\ {\rm cm^2\,g^{-1}}$ and
we assume the standard efficiency factor $\eta_{eff} = 0.1$
(e.g. Frank et al. 1992); and radii in Schwarzschild units
\[R = r R_{Schw},\ \ \ \ R_{Schw} = \frac{2 G M}{c^2} = 2.95 \times 10^5 m\ 
\mbox{cm}.\]

If we choose the values of $r$, $m$, $\dot m$, $\alpha$ and $\beta_m$, and 
assume specific values for $\beta$ and $T$, equations (\ref{2.1}),
(\ref{2.2}) and (\ref{2.3}) yield all the other disk parameters.

\subsection{Electron-Positron Pair Equilibrium}

Some of the hot solutions we calculate in this paper have temperatures
of up to $10^{10} - 10^{11}$ K (see Figure \ref{rad.all}[d]), and it 
therefore becomes necessary to include in our model the effects of 
relativistic pair production and annihilation.  Since a complete 
solution of the pair balance problem is beyond the scope of our 
calculation, we make some simplifying assumptions.
  
Svensson (1982, 1984) has explored the properties of pair equilibria
in a uniform relativistic plasma where photons are generated via pair
annihilation and bremsstrahlung.
He finds that for temperatures below a certain critical value $T_c$
there are two equilibrium branches available that are characterized by
the ratio $z$ of the pair number density, $n_+$, to the number density
of protons, $n_p = n_e - n_+$: (1) an optically thin low-$z$ branch
($z \ll 1$) where pair production is dominated by particle-particle
processes, and (2) a high-$z$ branch ($z \ge 1$) where the Thomson
scattering optical depth is of order unity and photon-photon pair
production dominates.  We confine our calculations to the low-$z$
branch where only particle-particle processes are important.  Svensson
has shown that the electron-electron pair production rate is always
larger than the electron-proton rate by a factor of $\sim 10$.  This
allows us to ignore the contribution from electron-proton collisions.

With the above simplifications, we can solve for the equilibrium pair
density, $z$, analytically at any given temperature.  For a thermal
distribution of electrons and positrons, the pair annihilation rate
per unit volume is (Svensson 1982):
\begin{equation}
\label{2.4}
(\dot{n}_+)_{ann} = \pi c r_e^2 n_e n_+ g (\theta),
\end{equation}
where $\theta= k_B T/m_e c^2$ is the dimensionless electron
temperature, $r_e = e^2/m c^2$, and $g (\theta)$ is given by:
\begin{equation}
\label{2.5}
g (\theta) = \left[1 + \frac{2 \theta^2}{\ln{(1.12 \theta + 1.3)}}\right]^{-1}.
\end{equation}
For the electron-electron pair production rate we adopt the expression
used by White \& Lightman (1989):
\begin{equation}
\label{2.6}
(\dot{n}_+)_{ee} = c r_e^2 n_e^2 \left\{ \begin{array} {ll}
	2 \times 10^{-4}\ \theta^{3/2} \exp{(-2/\theta)}\ (1 + 0.015 \theta),
	& \mbox{if}\ \ \theta \ll 1, \\
	(112/27 \pi) \ \alpha_f^2\ (\ln{\theta})^3\ (1+0.058/\theta)^{-1},
	& \mbox{if}\ \ \theta \gg 1,
	\end{array}\right.
\end{equation}
where $\alpha_f$ is the fine structure constant.  Local equilibrium
requires that $(\dot{n}_+)_{ee}=(\dot{n}_+)_{ann}$, which allows us to
solve for $z$ as a function of the plasma temperature.  We find that
for the highest temperature of interest, $T = 10^{11}$ K, $z$ has a
maximum value $\sim 0.1$, and that it decreases monotonically in
cooler plasmas.  This implies that our low-$z$ assumption is
consistent.

Given $z$ we can compute the total number density of electrons in the
disk, $n_e = n_p (1+z)$, and number density of electrons plus
positrons, $n_{\pm} = n_p (1+2 z)$.
    
\subsection{Cooling Processes}

Since ions are considerably more massive than electrons and positrons,
viscous heating will affect primarily the ions (SLE, Phinney 1981,
Rees et al. 1982), which will then transfer this energy to the
electrons.  In their paper NY assume that the energy transfer occurs
only via Coulomb interactions, ignoring possible non-thermal coupling
mechanisms (e.g. Begelman \& Chiueh 1988).  Since Coulomb interactions
are not efficient at low densities, the ions are always hotter than
the electrons.  In this paper we assume that the ions and electrons
have been brought into thermal equilibrium by some process, without
explicitly specifying the actual coupling mechanism.

The cooling of the hot plasma occurs mainly through the electrons and
positrons.  In our model we consider three cooling processes:
bremsstrahlung, synchrotron, and Compton cooling by both
bremsstrahlung and synchrotron photons.

\paragraph{Bremsstrahlung Cooling}
The free-free radiation in a plasma is produced through five different
types of interactions: electron-electron ($e^- e^-$), electron-ion
($e^-i$), positron-electron ($e^- e^+$), positron-positron ($e^+
e^+$), and positron-ion ($e^+ i$).  However, $e^-i$ and $e^+ i$
processes can be combined together since their corresponding cooling
rates are identical, and the same is true for $e^- e^-$ and $e^+ e^+$
processes.  Thus the total bremsstrahlung cooling rate per unit volume
is just the sum of the rates for $e^{\pm} i$, $e^{\pm} e^{\pm}$, and
$e^- e^+$ processes; and we can write $q_{br}^- = q_{e i}^- + q_{e
e}^- + q_{+ -}^-$, provided we replace $n_p n_e$ by $n_p n_{\pm}$ and
$n_e^2$ by $(n_e^2 + n_+^2)$ in our calculations of $q_{e i}^-$ and
$q_{e e}^-$ respectively.

Following NY, we adopt the expressions from Stepney \& Guilbert (1983)
and Svensson (1982) for the $e^{\pm} i$, $e^{\pm} e^{\pm}$, and $e^-
e^+$ bremsstrahlung cooling rates:
\begin{eqnarray}
\label{2.8}
q^-_{e i} &=& 1.48\times 10^{-22}\ n_p n_{\pm}\ F_{e i}
    	\ \ {\rm erg\,cm^{-3}\,s^{-1}}; \\
\label{2.7}
q^-_{e e} & = & \left\{ \begin{array} {ll}
	   2.56\times 10^{-22} (n_e^2+n_+^2) \theta^{1.5} (1+1.1 \theta+
	   \theta^2-1.25 \theta^{2.5})\ {\rm erg\,cm^{-3}\,s^{-1}}, & 
	       \mbox{if}\ \ \theta < 1, \\
	   3.42\times 10^{-22} (n_e^2+n_+^2) \theta\ (\ln{(1.123 \theta)}+
	   1.28)\ \ {\rm erg\,cm^{-3}\,s^{-1}}, &
	       \mbox{if}\ \ \theta \ge 1; 
                   \end{array}\right. \\	
\label{2.7a} 
q^-_{+ -} & = & \left\{ \begin{array} {ll}
	   3.43\times 10^{-22}\ n_e n_+\ (\theta^{0.5}+1.7 \theta^2)
	       \ \ {\rm erg\,cm^{-3}\,s^{-1}}, & 
	       \mbox{if}\ \ \theta < 1, \\
	   6.84\times 10^{-22}\ n_e n_+\ \theta\ (\ln{(1.123 \theta)}+1.24)
               \ \ {\rm erg\,cm^{-3}\,s^{-1}}, &
	       \mbox{if}\ \ \theta \ge 1; 
                   \end{array}\right. 
\end{eqnarray}
where 
\begin{equation}
\label{2.9}
F_{e i} = \left\{ \begin{array} {ll}
	   4 \sqrt{\frac{2 \theta}{\pi^3}}\ (1+1.781\ \theta^{1.34}), &
	       \mbox{if}\ \ \theta < 1, \\
	   \frac{9 \theta}{2 \pi}\ (\ln{(1.123 \theta +0.48)} +1.5), &
	       \mbox{if}\ \ \theta \ge 1.
	          \end{array} \right. 
\end{equation}

\paragraph{Synchrotron Cooling}
Synchrotron cooling is very important in our model, both because we
assume an equipartition magnetic field in the disk and because we
consider relativistic temperatures for the electrons and positrons.

Pacholczyk (1970) has calculated the spectrum of synchrotron emission
by a relativistic Maxwellian distribution to be
\begin{equation} 
\label{2.10}
\epsilon_{s}\ d \nu = 4.43 \times 10^{-30}\ \frac{4 \pi \nu n_{\pm}}
{K_2(1/\theta)}
\ I\left(\frac{x_M}{\sin{\phi}}\right)\ d \nu\ {\rm erg\,cm^{-3}\,s^{-1}},
\end{equation}
where 
\begin{eqnarray}
x_M = \frac{2 \nu}{3 \nu_0 \theta^2},\ \ \ \ \nu_0 = \frac{e B}{2 \pi m_e c}, 
\nonumber
\end{eqnarray}
$\phi$ is the angle between the velocity vector of the electrons and
the direction of the local magnetic field, and $I(x)$ is a tabulated
function.  Positrons produce synchrotron radiation at the same rate as
electrons do and therefore in equation (\ref{2.10}) we have used the
total number density, $n_{\pm}$, instead of $n_e$.

For an isotropic velocity distribution we can average
$I(x_M/\sin{\phi})$ over $\phi$ to get a new function
$I^{\prime}(x_M)$ for which a fitting function was found by Mahadevan,
Narayan \& Yi (1996):
\begin{equation}
\label{2.11}
I^{\prime}(x_M)=\frac{4.0505}{x_M^{1/6}} \left(1+ \frac{0.40}{x_M^{1/4}} +
\frac{0.5316}{x_M^{1/2}}\right)\ \exp\left(-1.8899 x_M^{1/3}\right).
\end{equation}
We substitute this function for $I(x_M/\sin{\phi})$ in equation
(\ref{2.10}).

Equations (\ref{2.10}) and (\ref{2.11}) are valid only for optically
thin emission. However, below some critical frequency $\nu_c$ the
emission becomes self-absorbed and that has to be taken into account
in computing the total cooling rate.  We estimate $\nu_c$ as the 
frequency at which the synchrotron emission from a thin annulus
of height $2 H$, radius $R$, and thickness $\Delta R$ is equal to the 
blackbody emission (in the Rayleigh-Jeans limit) from the upper and 
lower surfaces of the annulus (NY presented a similar argument but they  
used a spherical rather than cylindrical geometry).  This condition gives 
us the equation:
\begin{equation}
\label{2.12}
2 H \cdot 2 \pi R \Delta R \cdot \epsilon_s\ d \nu = 2 \cdot 2 \pi R \Delta R 
\cdot 2 \pi \frac{\nu^2_c}{c^2} k T\ d \nu,
\end{equation}
which we can solve numerically for given values of $R$, $H$, and $T$
to obtain $\nu_c$.

Given the value of the critical frequency, we can estimate the total
synchrotron emission per unit volume as follows.  We assume that for
frequencies below $\nu_c$ the emission is completely self-absorbed so
that the volume emissivity can be approximated by the blackbody
emission from the surface of the disk divided by the disk volume.
Above $\nu_c$ the emission is optically thin and equation (\ref{2.10})
can be used.  To get the total cooling per unit volume by synchrotron
emission, we then integrate over frequency:
\begin{eqnarray}
\label{2.13}
q^-_s & = & \frac{2 \pi R^2}{2 H \pi R^2}\ \int_0^{\nu_c}{2 \pi \frac{\nu^2}
{c^2}k T\ d \nu} +  \int_{\nu_c}^{\infty}{\epsilon_{\nu}\ d \nu} \nonumber \\
 & = & \frac{2 \pi k T \nu_c^3}{3 H c^2} + 
6.76\times 10^{-28}\ \frac{n_{\pm}}{K_2 (1/\theta) a^{1/6}} \left[
\frac{1}{a_4^{11/2}} \Gamma \left(\frac{11}{2}, a_4 \nu_c^{1/3}\right)
\right. + \nonumber \\
 &+& \left. \frac{a_2}{a_4^{19/4}} \Gamma \left(\frac{19}{4}, a_4 
\nu_c^{1/3}\right) +
\frac{a_3}{a_4^4}\ \left(a_4^3 \nu_c + 3 a_4^2 \nu_c^{2/3} + 
6 a_4 \nu_c^{1/3} +6\right)\ e^{-a_4 \nu_c^{1/3}} \right],
\end{eqnarray}
where the parameters $a_1$, $a_2$, $a_3$, $a_4$ are defined as
\begin{equation}
\label{2.14}
a_1 = \frac{2}{3 \nu_0 \theta^2},\ \ \ a_2=\frac{0.4}{a_1^{1/4}},\ \ \ 
a_3 = \frac{0.5316}{a_1^{1/2}},\ \ \ a_4 = 1.8899 a_1^{1/3}.
\end{equation}

\paragraph{Compton Cooling}
Since our solutions are hot, Comptonization of soft photons by hot
electrons becomes an important cooling mechanism.  It is an especially
important process for higher values of the accretion rate, when the
scattering depth is of order unity and synchrotron emission is the
dominant photon production mechanism, i.e. most photons are initially
very soft.

We compute the Compton cooling rate using the Comptonized energy
enhancement factor $\eta$, which is defined as the average ratio of
the energy of a photon at escape to its initial energy.  We adopt the
following prescription for $\eta$, which is derived in Appendix A:
\begin{equation}
\label{2.15}
\eta = e^{s (A-1)} \left[1-P (j_m+1, A s)\right] + \eta_{max} P (j_m+1, s),
\end{equation} 
where $P (a, x) = [1/\Gamma(a)] \int_0^x{t^{a-1} e^{-t}\ d t}$ is the
incomplete gamma function and
\begin{eqnarray}
\label{2.16}
A & = & 1+4 \theta + 16 \theta^2, \nonumber \\
s & = & \tau_{es} + \tau_{es}^2, \nonumber \\
j_m & = & \ln{(\eta_{max})}/\ln{(A)}, \nonumber \\
\eta_{max} & = & \frac{3 k T}{h \nu}.
\end{eqnarray}
The parameter $A$ is the average increase in energy of a soft photon
per scattering for a Maxwellian distribution of electrons (and
positrons) at temperature $\theta$, $\tau_{es} = 2 n_{\pm} \sigma_T H$
is the Thomson optical depth (counting both electrons and positrons),
$j_m$ is the number of scatterings required for the maximum possible
energy enhancement $\eta_{max}$, and $\nu$ is the initial photon
frequency.  By definition, the Comptonized flux is the original flux
multiplied by $\eta$.

Since the synchrotron spectrum is strongly peaked at the critical
frequency $\nu_c$, we estimate the cooling rate due to the
Comptonization of the synchrotron radiation simply by computing $\eta$
for $\nu = \nu_c$:
\begin{equation}
\label{2.17}
q^-_{C,s} = \eta (\nu_c)\ q^-_s. 
\end{equation}

The spectrum of bremsstrahlung radiation is practically flat for
frequencies between the bremsstrahlung self-absorption frequency,
$\nu_{br}$, and the exponential cutoff at $\nu = k T/h$. The emission
at frequencies $\nu < \nu_{br}$ is negligible compared with the
integrated bremsstrahlung cooling rate.  Therefore, we can approximate
free-free emission per unit frequency as the ratio $q^-_{br}/(\frac{k
T}{h} - \nu_{br})$.  We estimate $\nu_{br}$ as the frequency at which
the free-free absorption opacity, $\kappa_{\nu}^{ff}$ (which is
frequency dependent) is equal to the electron scattering opacity,
$\kappa_{es}$.  Below $\nu_{br}$ Comptonization is not important,
since absorption dominates over scattering.  Therefore to calculate
the Comptonization enhancement we perform the following numerical
integration:
\begin{equation}
\label{2.18}
q^-_{C,br} = \int_{\nu_{br}}^{\frac{k T}{h}}{\eta (\nu)\ \frac{ d q^-_{br}}
{d \nu}\ d \nu} = \int_{\nu_{br}}^{\frac{k T}{h}}{\eta (\nu)\ \frac{q^-_{br}}
{\frac{k T}{h} - \nu_{br}}\ d \nu}.
\end{equation}

\paragraph{Cooling of the Optically Thick Gas}

The total cooling rate for the optically thin gas is simply the sum of
the two components calculated above:
\begin{equation}
\label{2.19}
q^- = q^-_{C,br} + q^-_{C,s}.
\end{equation}
This rate is appropriate for the hot solutions, which are generally
optically thin.  However, to reproduce the equilibrium solution
corresponding to the cool optically thick Shakura \& Sunyaev disk, we use
the generalized cooling formula given by NY:
\begin{equation}
\label{2.20}
q^- = \frac{4 \sigma T^4/H}{1.5 \tau + \sqrt{3} + 
\frac{4 \sigma T^4}{H} \left(q^-_{C,br} + q^-_{C,s}\right)^{-1}},
\end{equation}
where $\tau = \tau_{es} + \tau_{abs}$ is the total optical depth of
the disk in the vertical direction and $\tau_{abs} = (H/4 \sigma_B
T^4)\ \left( q^-_{C,br} + q^-_{C,s}\right)$ is the optical depth for
absorption.  For a small optical depth, equation (\ref{2.20}) reduces
to equation (\ref{2.19}), and therefore this formula can be used in
the limit of both very high and very low optical depth.

\paragraph{Radiation Pressure}

The total cooling rate allows us to compute the radiation pressure in
the disk,
\begin{equation}
\label{2.21}
p_r = \frac{q^- H}{2 c} (\tau + \frac{2}{\sqrt{3}}).
\end{equation}
The gas pressure in the accreting flow is simply:
\begin{equation}
\label{2.22}
p_g = \rho k T \left(\frac{1}{\mu_i m_u} + \frac{1}{\mu_e m_u} \right),
\end{equation}
where $\mu_i$ and $\mu_e$ are the effective molecular weights of the
ions and electrons, given by
\begin{equation}
\label{2.23}
\mu_i = \frac{4}{1 + 3 X} = 1.23,\ \ \ \ \mu_e = \frac{2}{1 + X} = 1.14,
\end{equation}
with numerical values computed for a hydrogen mass fraction $X=0.75$
(NY).  Given $p_r$ and $p_g$, the parameter $\chi$, defined in \S2.1,
is readily computed:
\begin{equation}
\label{2.24}
\chi = \beta/\beta_m = p_g/(p_r+p_g).
\end{equation}

\section{Properties of the Solutions}

An equilibrium thermal state for the accretion flow requires that
viscous heating minus the advected energy is exactly balanced by the
total radiative cooling at each radius:
\begin{equation}
\label{3.1}
q^+ (1-f) = q^-.
\end{equation} 
For fixed values of $m$, $r$, $\dot{m}$, and $\beta_m$, the equations 
(\ref{2.1})-(\ref{3.1}) comprise a closed set which can be solved to 
obtain all the relevant parameters of the accretion flow.  We solve the 
equations numerically using $T$ and $\chi$ as free variables.  We begin 
by assuming arbitrary values for $T$ and $\chi$ and use equations
(\ref{2.2})-(\ref{2.6}) to calculate all the parameters of the plasma,
including the pair fraction.  Then we use equations
(\ref{2.8})-(\ref{2.23}) to calculate the cooling rate $q^-$ and the
gas and radiation pressures, $p_g$ and $p_r$.  Now we compute $\chi =
p_g/(p_r+p_g)$ (see equation (\ref{2.24})) and compare the result with
the initially assumed value.  If the two values are not equal, we vary
$\chi$ keeping $T$ fixed until there is agreement between the initial
and final $\chi$.  We then vary $T$, optimizing $\chi$ at each step,
until the energy equation (\ref{3.1}) is satisfied.  At this point we
have a self-consistent equilibrium solution.

All calculations presented below were done for an equipartition
magnetic field strength in the disk, $\beta_m=0.5$.  However, we have
found that models with $\beta_m=0.95$ (where magnetic pressure
constitutes only $5 \%$ of the total pressure) produce practically the
same results.

\begin{figure}
\includegraphics{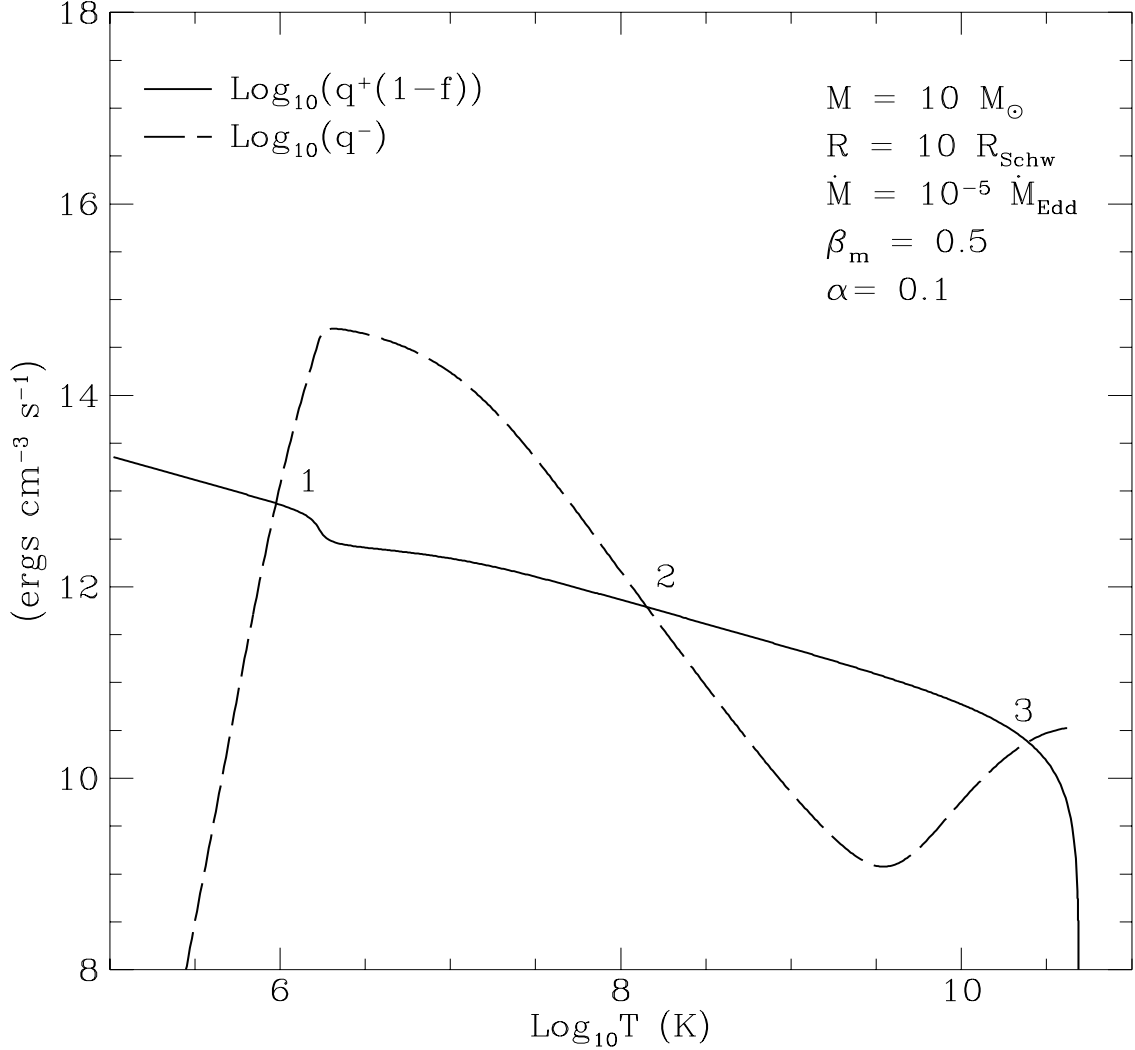}
\vskip 7.5in
\caption{\label{stab}
The rates of heating (solid line) and cooling (dashed line) of 
the accreting single-temperature gas plotted vs. gas temperature, $T$.  
Intersection points of the two curves mark
equilibrium states of the system.  Point 1 corresponds to the standard
thin disk solution, point 2 is equivalent to the unstable SLE
solution, and point 3 is our new, hot, thermally and viscously stable,
solution.}  
\end{figure}

Figure \ref{stab} illustrates the variations with $T$ of the two sides of
equation (\ref{3.1}); the solid line shows $q^+ (1-f)$ and the dashed
line shows the total radiative cooling $q^-$.  These results are for a
single-temperature accretion disk of mass accretion rate $\dot{M} =
10^{-5} \dot{M}_{Edd}$ around a black hole of mass $M = 10 M_{\odot}$
at radial distance $R = 10 R_{Schw}$; $\chi$ has been optimized at
each $T$.  The three points where the two curves intersect, labeled as
1, 2, and 3 on the figure, correspond to three equilibrium states of
the accretion disk where the energy balance equation (\ref{3.1}) is
satisfied.  The equilibrium point 1 corresponds to the standard
Shakura \& Sunyaev (1973) thin disk; the middle point is the
one-temperature equivalent of the unstable Shapiro, Lightman, \&
Eardley (1976) solution; the equilibrium point labeled 3 is a new
hot solution which which does not seem to have been discussed before 
in the literature.  It is related to the advection-dominated one-temperature
solution of Abramowicz et al. (1995), but with some important differences
as we argue later.

Note that the slope of the cooling curve in Figure \ref{stab} changes twice.
At $T \approx 10^{6.2}$ K the change of slope indicates a transition
from an optically thick to an optically thin flow. There is a
corresponding kink in the heating curve, since at this point the
cooling rate is at its maximum and radiation pressure briefly dominates 
over the gas pressure, causing a sharp decrease in gas density.  For 
$T < 10^{6.2}$ K, the gas cools as a blackbody and the cooling rate 
increases with increasing temperature.  For $T > 10^{6.2}$ K, the
dominant cooling process is optically thin bremsstrahlung, which is
less efficient at higher temperatures.  At $T \approx 10^{9.5}$ K the
slope of $\log{(q^-)}$ changes again, indicating a point where
synchrotron radiation becomes the dominant cooling process.  In
contrast to bremsstrahlung, the synchrotron emission increases with
increasing temperature, hence the positive slope of the cooling curve
at these high temperatures.

By inspection it is obvious that equilibria 1 and 3 are thermally
stable.  In both of these equilibria, if the gas is perturbed to a
higher temperature, the rate of cooling becomes greater than the rate
of heating, allowing the medium to cool back to the equilibrium value of
$T$.  Similarly, if $T$ becomes slightly smaller, the heating rate
exceeds the cooling rate and the gas heats up back to the equilibrium
temperature.  The equilibrium 2 on the other hand is unstable.  Here a
small increase or decrease in $T$ leads to a runaway situation where
$T$ deviates progressively more rapidly from the equilibrium value.

\begin{figure}
\includegraphics{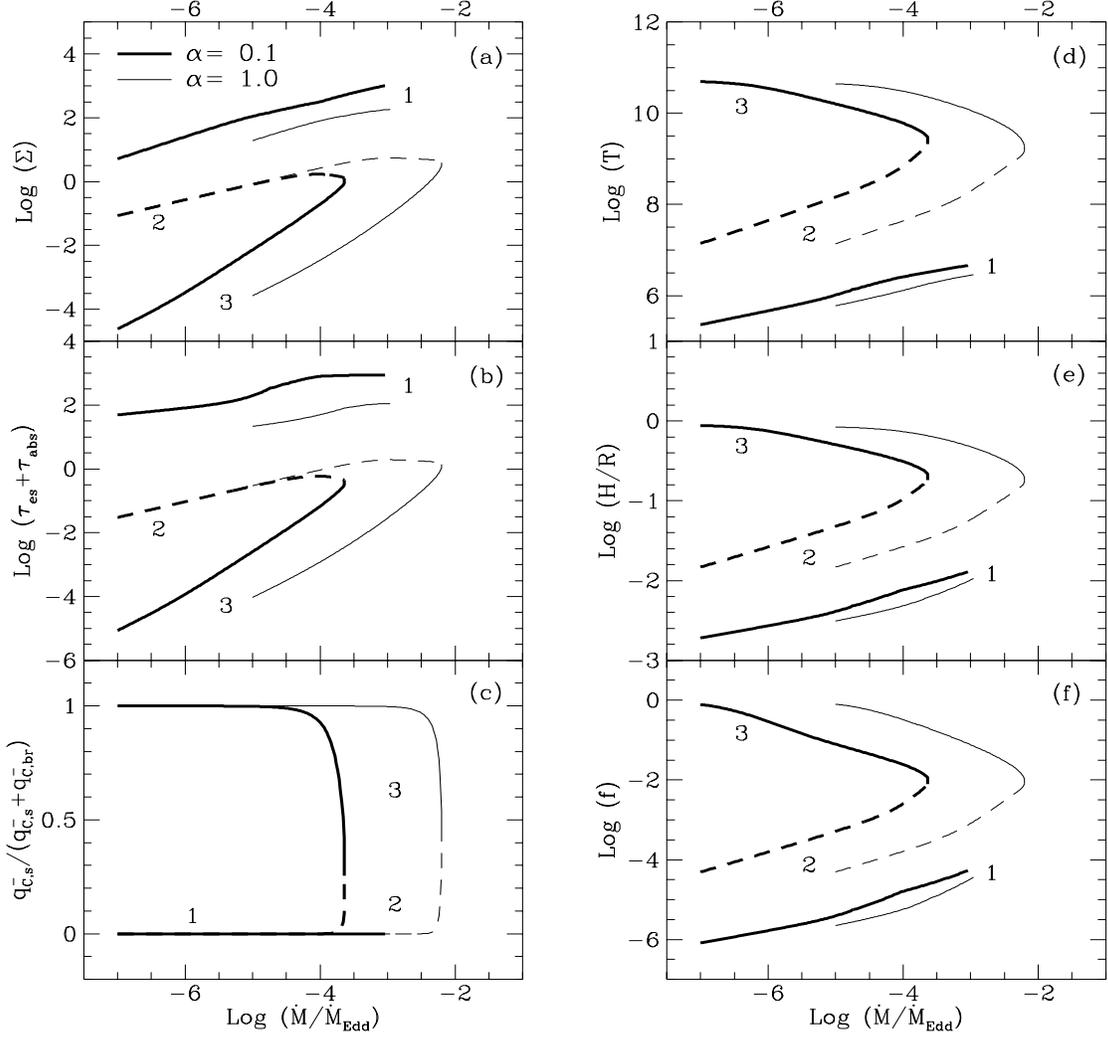}
\vskip 6.5in
\caption{\label{mdot.alpha}
Thermal equilibria of single-temperature accretion disks
with $\beta_m = 0.5$ around a $10 M_{\odot}$ black hole at a radius of
$R = 10 R_{Schw}$.  The heavy and thin lines represent solutions with 
$\alpha = 1$ and $\alpha = 0.1$ respectively.  The three branches
labeled as 1, 2, and 3 correspond respectively to the standard thin
disk, unstable SLE disk (indicated by a dashed line), and our hot and
stable solution (as in Figure 1).  In the six panels we plot (a) 
surface
density, $\Sigma$ (${\rm g\,cm^{-2}}$), (b) total optical depth, 
$(\tau_{es} + \tau_{abs})$, (c) fraction of the total cooling due to 
Comptonized synchrotron radiation, (d) temperature, $T$ (K), (e) the 
vertical scale height, $H/R$, and (f) advected energy fraction, $f$,
all as functions of the accretion rate $\dot{M}$.  Note that the hot 
solution branch (labeled 3) is optically thin, has essentially 
spherical geometry ($H/R \approx 1$), and cools mainly by synchrotron 
radiation.}
\end{figure}

We have computed the properties of the accreting gas corresponding to
the three equilibria for different values of $\dot{M}$.  The resulting
curves for a central black hole of mass $M = 10 M_{\odot}$ at fixed
radial distance $R = 10 R_{Schw}$ are plotted in Figure \ref{mdot.alpha} 
for $\alpha = 0.1$ (heavy line) and $\alpha = 1.0$ (thin line).  The 
curves are labeled 1, 2, and 3 to indicate correspondence with the 
equilibria labeled 1, 2, 3 in Figure \ref{stab}.  The six panels in 
Figure \ref{mdot.alpha} show how
various physical properties of the solutions depend on the accretion
rate.  The unstable SLE solution (branch 2) is indicated by a dashed
line.

For $\alpha = 0.1$ the topology of our results is qualitatively
similar to that of the corresponding solutions for two-temperature
disks described by Chen et al. (1995).  We see that the two hotter
branches merge together at some critical accretion rate
($\dot{M}_{crit}$), above which the cooling in the disk is so
efficient that the only equilibrium allowed is the standard thin
disk solution.  For $\alpha = 1$ the hot branch extends to much higher
values of the accretion rate; in fact, the value of $\dot{M}_{crit}$
goes up by $\sim 2$ orders of magnitude, in agreement with the results
of Abramowicz et al. (1995) and NY.  Somewhat surprisingly, the topology
of the solution curves for $\alpha=1$ continues to be similar to that
for $\alpha = 0.1$.  This is different from the result reported by
Chen et al. (1995), who found that for $\alpha = 1$ the middle branch
merged with the cold disk solution, whereas the hottest branch
extended independently upward to become the optically-thick
advection-dominated solution discovered by Abramowicz et al. (1988).
The value of the critical $\alpha$ at which the topology changes is,  
however, sensitive to the details of the
radiation processes included in the model (M. Abramowicz, private
communication).  In fact, we do find the same topology change in our results
when we push $\alpha$ above $\sim 1.7$.

For both $\alpha=0.1$ and $\alpha=1$, the curves corresponding to the
three equilibria in Figure \ref{mdot.alpha}(a) have positive slopes,
i.e. $\partial{\dot{M}}/\partial{\Sigma} > 0$.  This implies that all
three solutions are viscously stable (cf. Frank, King, \& Raine 1992,
p.103).  Note also that our new hot solution is optically thin for all
values of $\dot{M}$ where the solution exists (Figure \ref{mdot.alpha}[b]) 
and cools mainly through synchrotron and Comptonized synchrotron processes
(Figure \ref{mdot.alpha}[c]).

\begin{figure}
\includegraphics{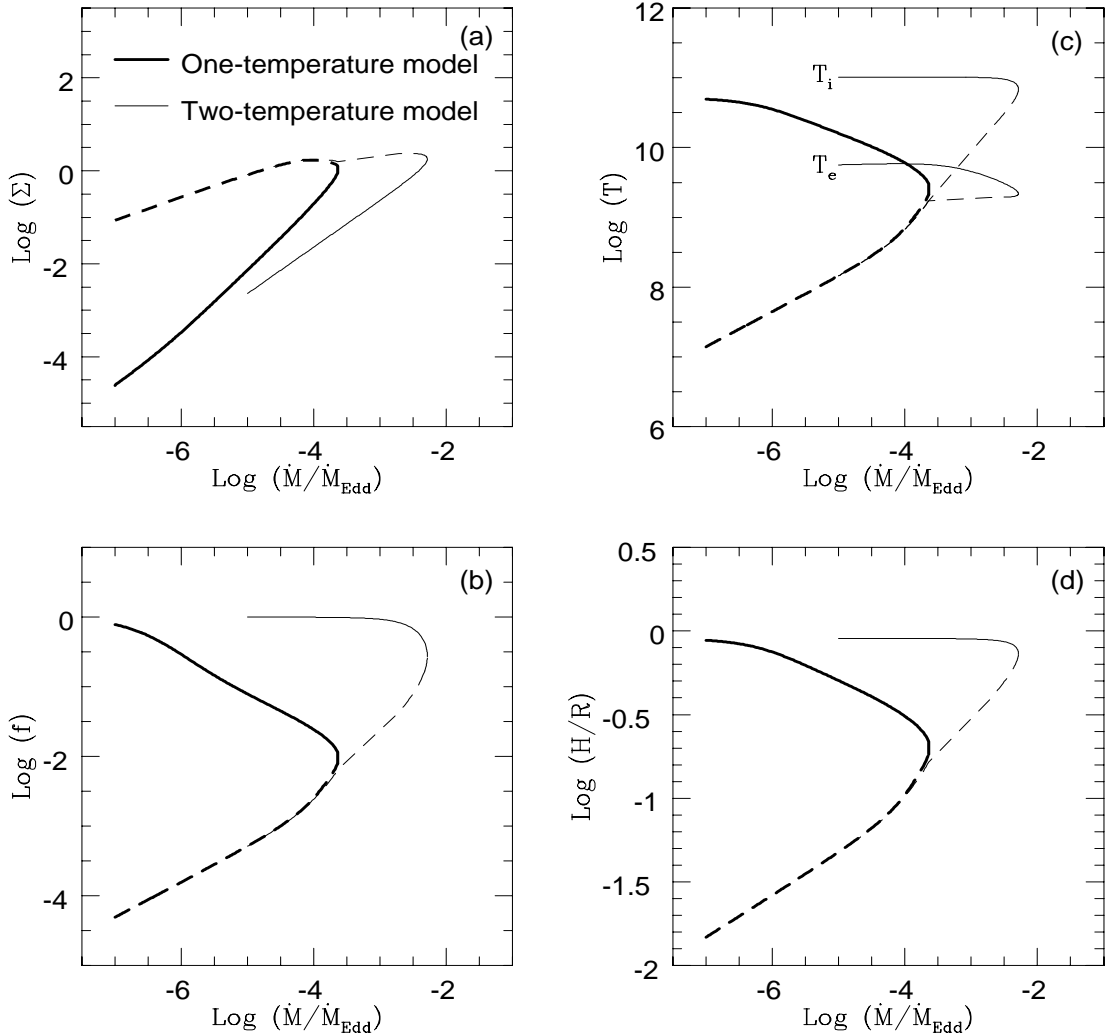}
\vskip 7.0in
\caption{\label{mdot.2temp}
Thermal equilibria of single-temperature (heavy line) and
two-temperature (thin line) accretion disks with $\beta_m = 0.5$ and
$\alpha = 0.1$ around a $10 M_{\odot}$ black hole at a radius of
$r=10$.  The thin disk branch is not shown.  The four panels show 
(a) surface density, $\Sigma$ (${\rm g\,cm^{-2}}$), (b) advection 
parameter, $f$, (c) temperature, $T$ (K), and (d) the scale height, 
$H/R$, plotted as functions of $\dot{M}$. The unstable branch is shown
by dashed lines.  On the stable branch, the single-temperature
solution is less advection-dominated and has a higher electron temperature
than the two-temperature solution.}
\end{figure}

Figure \ref{mdot.2temp} directly compares the two-temperature (thin line) 
and one-temperature (heavy line) equilibrium solutions for a disk with
$\alpha = 0.1$. 
For temperatures below $T \approx 10^9$ K the two curves merge, since 
the electron and ion temperatures in the two-temperature solution are 
nearly equal (Figure \ref{mdot.2temp}[c]) and therefore,
two-temperature and one-temperature solutions are effectively the
same.  However, for $T \gsim 10^9$ K the two solutions differ
significantly.  The most important distinction is in the value of the
advected energy fraction $f$ (Figure \ref{mdot.2temp}[b]).  For the 
hot two-temperature
solution, we have $f\rightarrow1$ at all $\dot M$, showing that this
branch is always advection-dominated.  In contrast, we see that our
one-temperature hot solution is cooling-dominated ($f < 0.1$) for a
wide range of accretion rates.  This is a completely new result, as
prior to this the only hot stable solutions known were all
advection-dominated (NY, Abramowicz et al. 1995, Chen 1995, Chen et
al. 1995).  Since the cooling rate is given by $q^- = q^+ (1-f)$, we
see that, at a given accretion rate, our single-temperature disk with
its low value of $f$ is considerably more luminous than the equivalent
two-temperature flow.

In the case of the two-temperature advection-dominated hot solutions 
discussed by NY and Abramowicz et al. (1995), the thermal stability of 
the solution to long-wavelength perturbations has been shown to be the 
direct result of advection.  However, our hot one-temperature solution 
branch 3 is cooling-dominated and has very little advection.
Why is this solution stable?  The answer is obvious from Figure \ref{stab},
where we see that the introduction of synchrotron cooling causes the
cooling curve $q^-(T)$ to be steeply positive for $T>10^{9.5}$ K.
Therefore, thermal stability is achieved in this solution because of 
the usual reason, namely that the cooling increases more rapidly than 
the heating.  Note that we would not have obtained this result if we had 
included only bremsstrahlung cooling.  In fact, Abramowicz et al. (1995)
considered the pure bremsstrahlung one-temperature case and they obtained 
a hot advection-dominated solution.  Thus, the inclusion of synchrotron 
cooling has an important effect on the basic physics of the accretion 
flow.  Although a full linear stability analysis
along the lines of Kato et al. (1995) has not been done on our new
solutions, we suspect that these solutions are stable to all linear
perturbations, regardless of wavelength.

\begin{figure}
\includegraphics{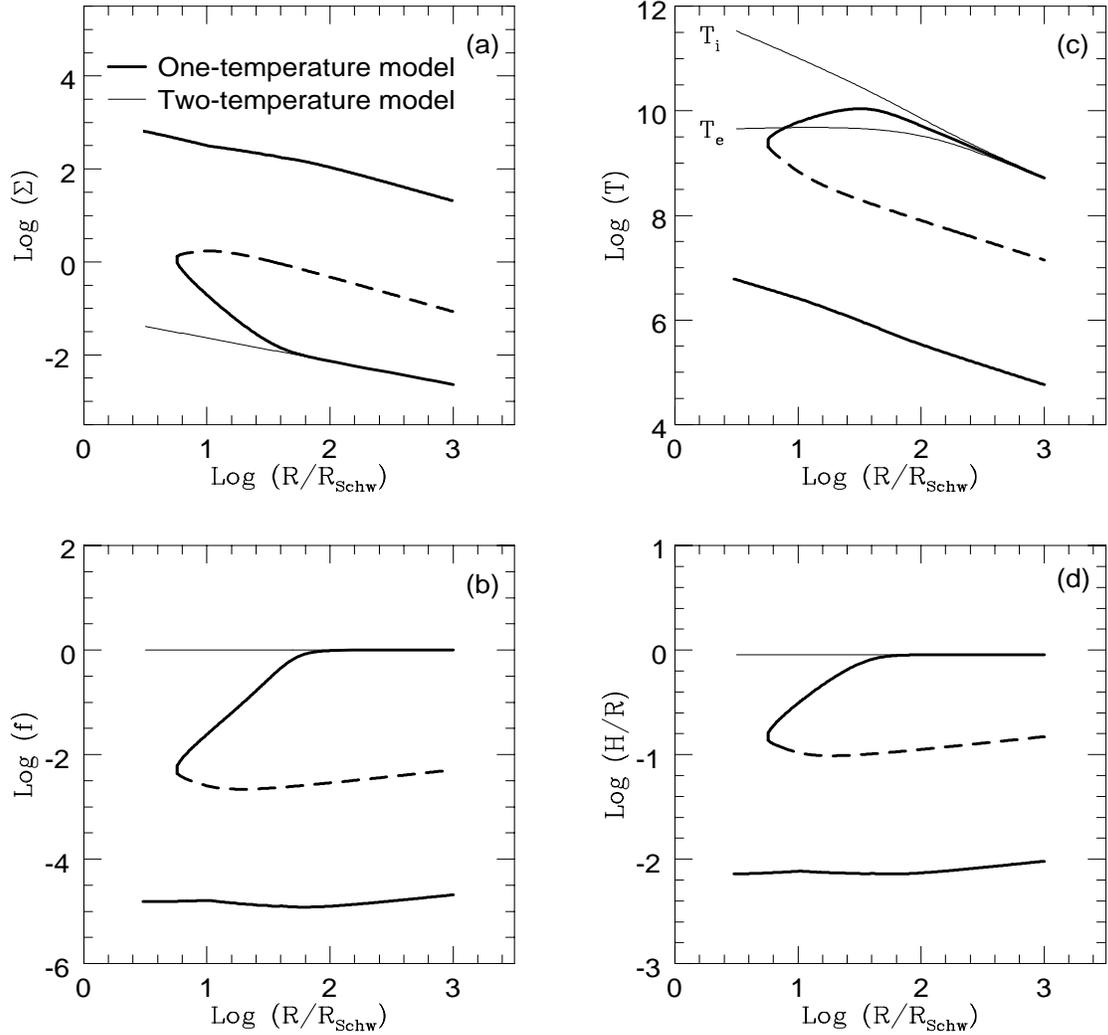}
\vskip 7.3in
\caption{\label{rad}
Radial profiles of thermal equilibria for a
single-temperature accretion disk with $\beta_m = 0.5$, $\alpha =
0.1$, and $\dot{M}/\dot{M}_{Edd} = 10^{-4}$ around a $10 M_{\odot}$
black hole (heavy line).  For comparison we also plot the hot 
stable branch of the two-temperature disk (thin line).  As expected,
the two models give identical results at large radii.  Closer to the black
hole, however, the single-temperature solution becomes
cooling-dominated while the two-temperature solution remains
advection-dominated.  Note the maximum in the disk temperature at $R
\approx 40 R_{Schw}$.}
\end{figure}

Figure \ref{rad} illustrates the radial structure of a disk with $\alpha =
0.1$ for a fixed value of the accretion rate, $\dot M=10^{-4}\dot
M_{Edd}$.  The inner edge was chosen to be at $R = 3 R_{Schw}$,
corresponding to the last stable orbit.  As expected, we see that the
two-temperature (thin line) and one-temperature (heavy line) solutions
merge at large radii where the temperature is low and even the
two-temperature plasma has effectively a single temperature (see
Figure \ref{rad}[c]).  Thus, away from the center, the flow is 
advection-dominated (see Figure \ref{rad}[b]) and almost spherical 
with $H/R \sim 1$ (Figure \ref{rad}[d]), 
regardless of whether we have a two-temperature or one-temperature 
solution.  However, closer to the accreting object, the flow
becomes increasingly cooling-dominated in the one-temperature case and
settles into a disk with $H/R \sim 0.15$ at the inner edge.  The
resulting geometry -- thinnish disk near the inner edge and
quasi-spherical flow at larger radii -- is unusual for accretion disk
models.  Another interesting feature of our one-temperature solution
is that the temperature profile of the disk is not monotonic but has a
well defined maximum at $R \sim 40 R_{Schw}$.  Therefore, the hottest
part of the spectrum does not come from the inner edge of the disk.

NY showed that their
advection-dominated hot two-temperature solution exists only for
accretion rates $\dot M$ less than a critical rate $\dot M_{crit}$,
which depends on $\alpha$ and $R$ (see also Abramowicz et al. 1995).
We find a similar result also in the case of our cooling-dominated hot
one-temperature solution.  Figure \ref{mcrit} shows the critical accretion rate
$\dot M_{crit}$ for our solution as a function of $R$ as well as the
variation of the gas temperature.  The two panels on the left (Figure
5[a,b]) compare the results for the one-temperature and two-temperature 
solutions for $\alpha = 0.1$.  We see that the electron temperatures 
for the two solutions are nearly equal, but the behavior of 
$\dot{M}_{crit}$ is quite different in the two cases.  For a two-temperature
flow, $\dot{M}_{crit}$ is roughly independent of $R$ for $R\lsim 10^3 R_{Schw}$
and is given approximately by $\dot{M} \sim 0.3 \alpha^2 \dot{M}_{Edd}$
(see NY and Abramowicz et al. 1995 for a discussion of the $\alpha^2$ 
scaling).  In contrast, $\dot{M}_{crit}$ decreases almost linearly with 
decreasing $R$ in the one-temperature case, so that $\dot{M} \sim 3 10^{-3} 
\alpha^2 (R/R_{Schw}) \dot{M}_{Edd}$ for $R \lsim 10^3 R_{Schw}$
Thus, near the 
black hole, $\dot{M}_{crit}$ for the one-temperature solution is smaller
than that of the two-temperature solution by as much as two orders of
magnitude. 

The steep decrease of $\dot M_{crit}$ with decreasing $R$ leads to an
interesting effect which is already seen in Figure \ref{rad}.  For a wide range
of $\dot M$, it is possible to have $\dot M<\dot M_{crit}$ at large
radii, but to have $\dot M>\dot M_{crit}$ at small radii, with a
transition at some critical radius $R_{min}$.  For the example shown
in Figure \ref{rad}, we have $R_{min}=5R_{Schw}$.  In this case, we can have a
hot one-temperature solution for $R>R_{min}$, but the solution
disappears at $R=R_{min}$.  Below $R_{min}$, the only equilibrium
solution available is the cool thin accretion disk.  (Note that, because
of the very low $\dot{M} \sim 10^{-4} \dot{M}_{Edd}$, the Lightman-Eardley 
instability does not affect the thin disk solution in this case.)  Thus, 
the example shown in Figure \ref{rad} has an extremely unusual structure.  For
$3R_{Schw}<R<5R_{Schw}$, there is a standard thin disk.  Then, for
$5R_{Schw}<R<40R_{Schw}$, we have a hot, but cooling-dominated
thickish disk.  Finally, for $R>40R_{Schw}$, we have a regular
advection-dominated quasi-spherical flow.  There are several
discussions in the literature of models where an outer thin disk makes
a transition to a hot inner flow (e.g. SLE, Wandel \& Liang 1991,
Narayan, McClintock, \& Yi 1996, Narayan 1996), but here we have an
example of a model which does the opposite, namely transforms from an
outer hot advection-dominated flow via a cooling-dominated hot flow to
an inner thin disk.  (See Melia 1994 for the closest example discussed
previously in the literature.)

Figure \ref{rad.all} shows how the properties of the flow change as the 
accretion rate is reduced.  There are some interesting effects.  First, 
with decreasing $\dot M$, the critical radius $R_{min}$ disappears so 
that there is no longer any reason for the gas to make a transition to a
cool disk on the inside.  As $\dot{M}$ decreases still farther, even the
cooling-dominated zone of the hot disk disappears, and advection becomes 
more important throughout the flow.  Consequently the disk thickens
considerably, and the maximum in the temperature profile at large
radius becomes less apparent.  For $\alpha=0.1$ and
$\dot{M}/\dot{M}_{Edd} < 10^{-8}$ the flow is advection-dominated and
nearly virial at all radii.

Figure \ref{rad.all}[c] shows the relative importance of Comptonized 
synchrotron cooling as a function of $R$ at various $\dot M$.  We see that
synchrotron and Comptonized synchrotron cooling is dominant in the inner 
regions, $R < 100 R_{Schw}$, regardless of the accretion rate.  In the 
outer disk, on the other hand, where the flow is cooler and more 
advection-dominated, the cooling is mostly due to bremsstrahlung emission.

Though one-temperature solutions are restricted to smaller values of
$\dot M$ than the corresponding two-temperature solutions,
nevertheless we should note that for any given $\dot{M}$ they are
significantly more luminous.  This is because these flows are
cooling-dominated at small radii where most of the radiation is
emitted.  Figure \ref{lumin}(a) illustrates this luminosity increase in 
the case of $\alpha = 0.3$.  Here we plot the total disk luminosity as a
function of $\dot{M}$ for one-temperature (solid line) and
two-temperature (dashed line) solutions.  Note that at $\dot{M} =
10^{-4} \dot{M}_{Edd}$ the one-temperature disk is $\sim 10^4$ times
more luminous than its two-temperature counterpart.  

Figure \ref{lumin}(b) shows how the total luminosity of our 
one-temperature solutions changes
with $\alpha$.  For a given value of $\dot{M}$, higher-$\alpha$
solutions are less luminous, since they are more advection-dominated
(see Figure \ref{mdot.alpha}[e]).  However, higher values of $\alpha$ 
allow the hot
solution to survive up to higher accretion rates (see Figure \ref{mcrit}[c])
with correspondingly higher luminosities.  Thus, the most luminous
system that we can model as a hot one-temperature disk extending
down to $R = 3 R_{Schw}$ requires $\alpha = 1.0$ and $\dot{M}=10^{-3}
\dot{M}_{Edd}$, and has a total luminosity $L = 10^{-2.8} L_{Edd}$.
We could in principle use higher values of $\dot{M}$, but then as 
discussed above, we will have to include a cool thin disk at small radii
$(R < R_{min}$).

\section{Discussion}

Most of the work done so far on modeling hot accretion disks has been
based on two-temperature flows, where the electrons are much cooler
than the ions (SLE; Rees et al. 1982; Melia \& Misra 1993; Narayan \&
Yi 1995b; Narayan et al. 1995, 1996).  However, if there exist
mechanisms of energy transfer between ions and electrons that are more
efficient than Coulomb scattering (Phinney 1981, Rees et al. 1982,
Begelman \& Chiueh 1988), the plasma might be well-coupled and
the ions and electrons may have the same temperature.  In this paper
we have investigated how the properties of the hot, stable
two-temperature solution described by NY and Chen et al. (1995) are 
altered if ions and electrons in the accreting plasma are constrained 
to have the same temperature.

We start from the set of equations developed by NY, adding radiation pressure
and particle-particle pair production processes, and improving some of
the radiative cooling calculations.  We solve these equations for
various choices of the viscosity parameter, $\alpha$, and magnetic
pressure parameter, $\beta_m$, explicitly setting the ion and electron
temperatures equal to each other.  The resulting three branches of
solutions which we obtain (Figure \ref{stab}) correspond to the standard thin
disk (Shakura \& Sunyaev 1973, Novikov \& Thorne 1973, Lynden-Bell \&
Pringle 1974), the one-temperature equivalent of the unstable hot
solution discovered by SLE, and finally a new thermally stable hot
solution whose properties we discuss in detail (\S3).  The new
solution, which is identified by label 3 in Figures 1 and 2, is the
one-temperature equivalent of the two-temperature advection-dominated
solutions discussed by NY (see also Abramowicz et al. 1995, Chen et
al. 1995).
 
For a viscosity parameter $\alpha=0.1$, we find that the topological
relationships among the three solution branches are similar to those
found by Chen et al. (1995) and NY for the two-temperature solutions.
Specifically, we see that the two hot branches merge at some maximum
value of the mass accretion rate, which we call $\dot{M}_{crit}$, such
that for $\dot{M} > \dot{M}_{crit}$ the only solution available to the
flow is the standard thin Shakura \& Sunyaev disk configuration.  
At higher $\alpha$, we confirm the result of Chen et al. (1995) that 
the topology changes.  However, the critical $\alpha$ where the change 
occurs is 1.7 (at $R = 10 R_{Schw}$) for the one-temperature case we 
consider, rather then 0.3 in Chen et al. (1995).

\begin{figure}
\includegraphics{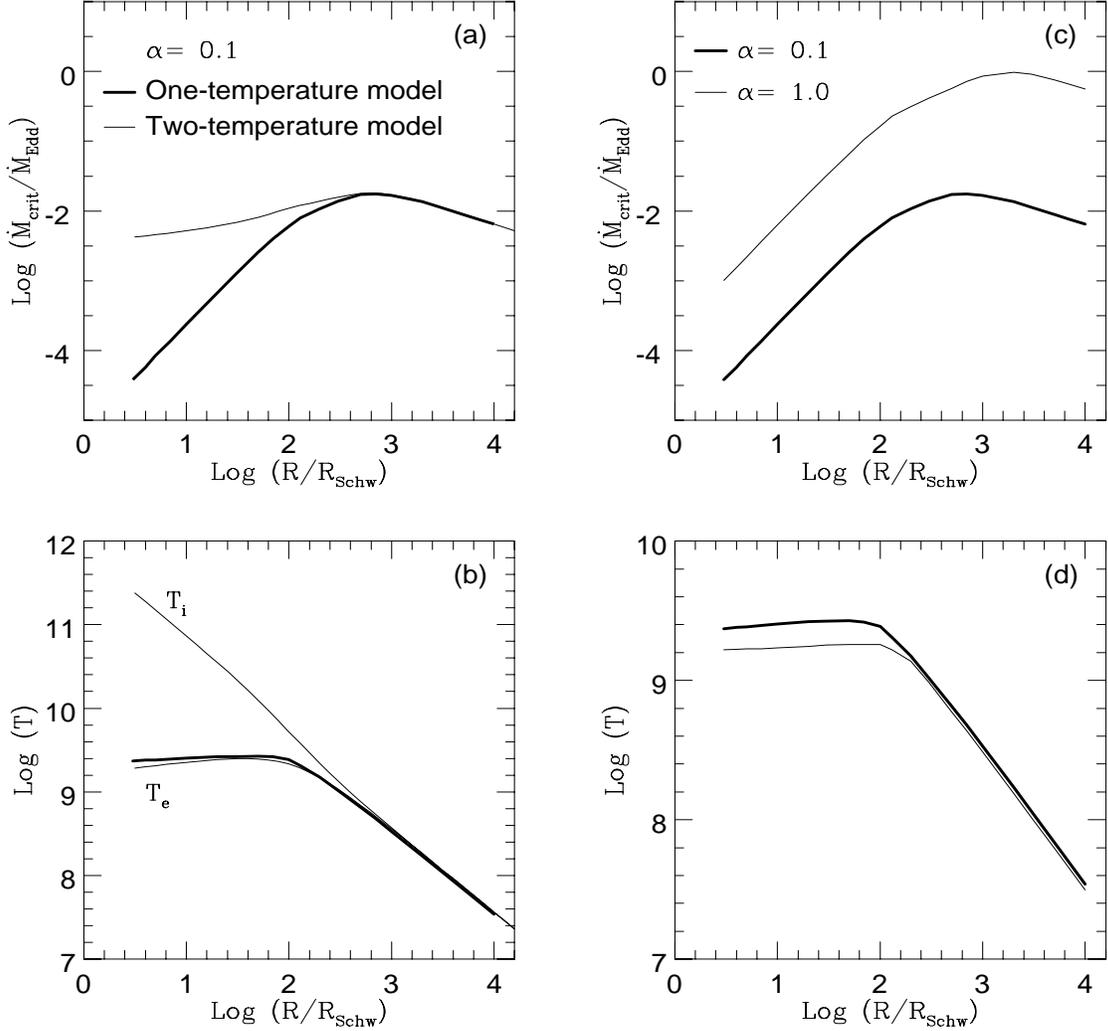}
\vskip 7.0in
\caption{\label{mcrit}
Critical accretion rate $\dot{M}_{crit}$ (upper panels) and 
corresponding temperature (lower panels) are plotted as functions of 
the radius. Panels (a) and (b) compare the results for the one- and
two-temperature models for $\beta_m = 0.5$ and $\alpha =0.1$.  The critical
$\dot{M}$ differs significantly in the two models, but the gas 
temperature in the single-temperature model is
practically always equal to $T_e$ in the two-temperature model.
Panels (c) and (d) show the results for a single-temperature disk
with $\beta_m = 0.5$ and two different values of $\alpha$.  Note that
although $\alpha=1.0$ allows models with values of $\dot{M}_{crit}$
higher by two orders of magnitude, the corresponding temperatures do
not depend strongly on $\alpha$.}
\end{figure}

\begin{figure}
\includegraphics{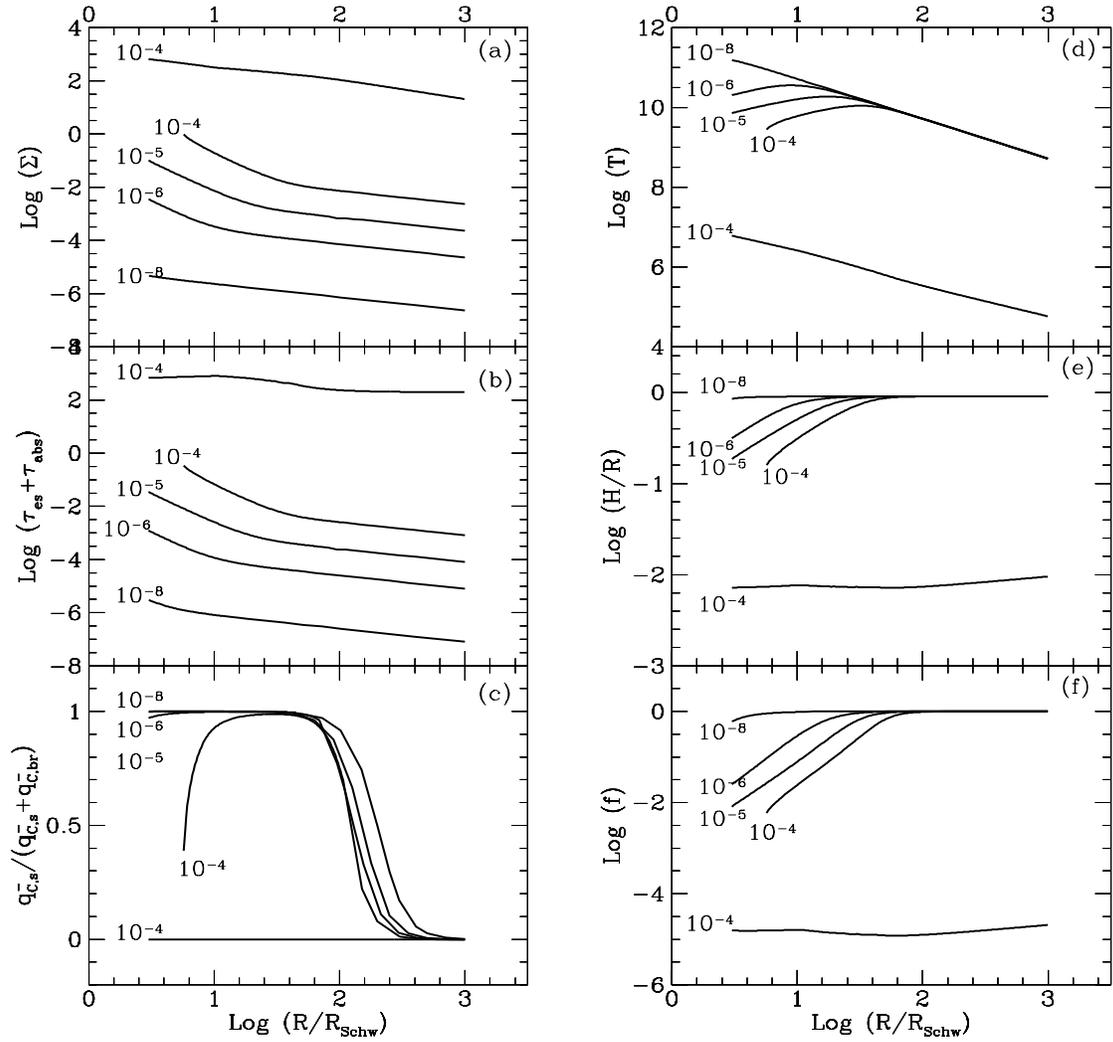}
\vskip 7.4in
\caption{\label{rad.all}
Radial profiles of hot single-temperature solutions, labeled by
the values of the accretion rate in units of $\dot{M}_{Edd}$.  The other 
parameters have the same values as in Figure 4.  Note that for 
$\dot{M} =
10^{-4} \dot{M}_{Edd}$ the hot branch does not extend all the way to
the inner edge of the disk at $R = 3 R_{Schw}$.  Therefore, the only
stable configuration at very small radii is the thin disk solution
which is also shown on the figure.}
\end{figure}

\begin{figure}
\includegraphics{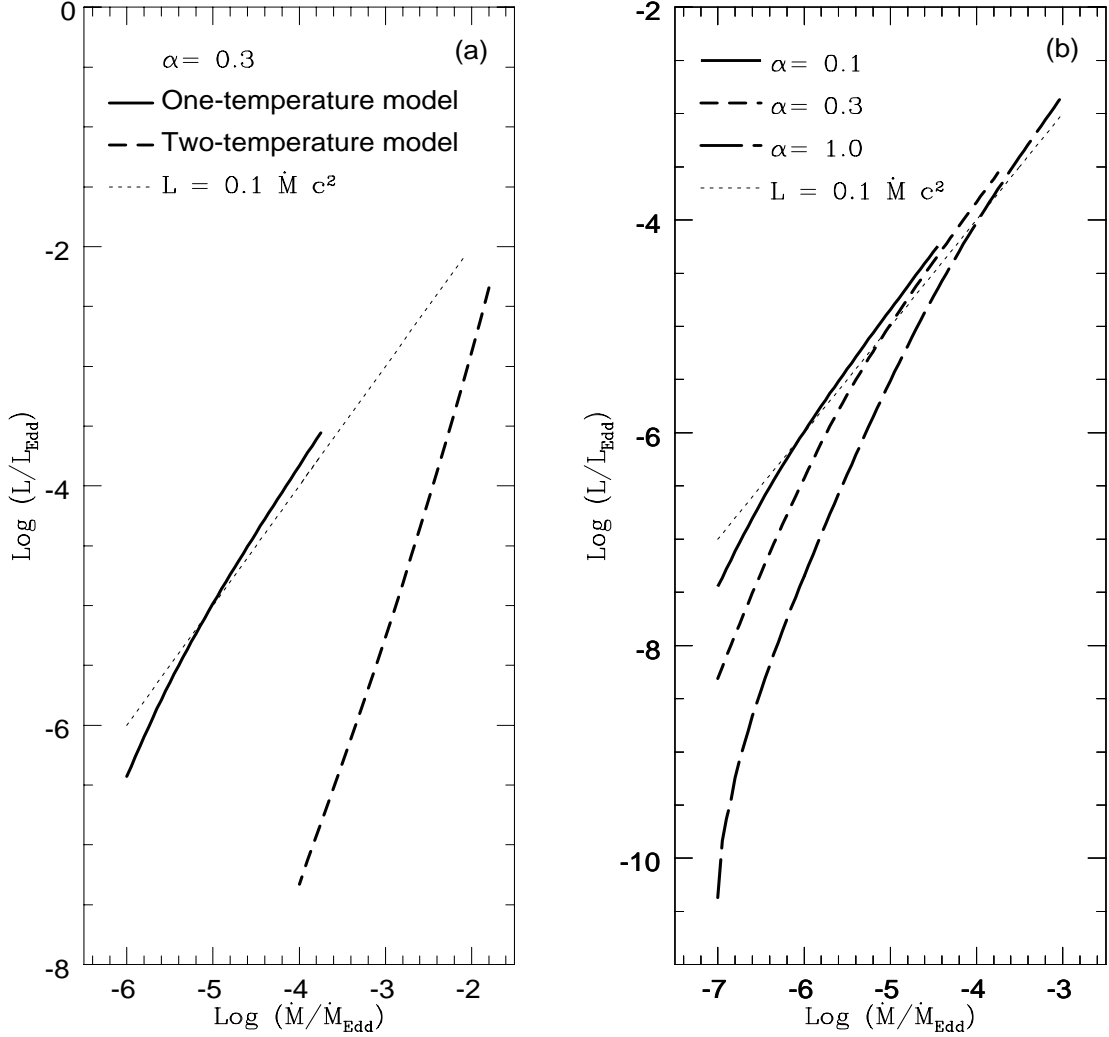}
\vskip 6.5in
\caption{\label{lumin}
(a) Compares the integrated luminosities for one-
and two-temperature hot accretion disks with $\beta_m = 0.5$, $\alpha=
0.3$, and $M = 10 M_{\odot}$.  The single-temperature disk is
clearly more efficient than a two-temperature disk for a given
$\dot{M}$ since its interior regions are cooling-dominated.  However,
the limiting $\dot M_{crit}$ of the one-temperature solution is lower 
than that of the two-temperature flow by two orders of magnitude.  
(b) Integrated luminosity of a hot single-temperature
accretion disk around a $10 M_{\odot}$ black hole as a function of the
mass accretion rate, for $\beta_m = 0.5$ and three different values of
$\alpha$.  Each curve was calculated up to the maximum value of the
accretion rate given by $\dot{M}_{max}= \dot{M}_{crit}(R = 3 R_{Schw})$.  The
dotted line shows the dependence expected if the radiative efficiency
is $10\%$.  Note that disks with lower values of $\alpha$ are more
luminous at a given $\dot M$, but have a lower limiting accretion rate
$\dot M_{crit}$.}
\end{figure}

We have computed the value of the critical mass accretion rate
$\dot{M}_{crit}$ as a function of radius for different values of
$\alpha$ (Figure \ref{mcrit}[c]) and $\beta_m$.  We find that $\dot{M}_{crit}$
varies roughly as $\alpha^2$ (Abramowicz et al. 1995, NY) and is 
relatively independent of $\beta_m$.  A new result is that $\dot{M}_{crit}$
varies almost linearly with $R$ for $R \lsim R_{Schw}$, since in a 
two-temperature flow $\dot{M}_{crit}$ is roughly independent of $R$ in that
region.  The maximum value 
of the accretion rate at which the hot solution exists near the inner
disk radius ($R = 3 R_{Schw}$) is $\dot{M}_{crit} \sim 3 \times 10^{-3} 
\alpha^2 \dot{M}_{Edd}$.  In comparison, two-temperature hot solutions have 
$\dot{M}_{crit} \sim 0.3 \alpha^2 \dot{M}_{Edd}$.

Though our single-temperature solutions are limited to lower mass
accretion rates than those allowed by two-temperature models, this is
compensated to some degree by strong cooling in the inner regions,
where $f < 0.1$ for $\dot{M} > 10^{-3} \dot{M}_{crit}(3R_{Schw})$ 
(Figure \ref{mcrit}[f]).
As a result, over the range of mass accretion rates, $10^{-3} <
\dot{M}/\dot{M}_{crit}(3R_{Schw}) < 1$, our solutions are significantly more
luminous than the corresponding two-temperature solutions (Figure
7[b]).  

This work represents the first example of a hot thermally stable
accretion solution which is also cooling-dominated and therefore an
efficient radiator.  Because advection plays a very minor role in our
new solutions, the thermal stability is not the result of advection as
in the equivalent two-temperature solutions (NY, Abramowicz et
al. 1995) or in the pure bremsstrahlung one-temperature solutions 
(Abramowicz et al. 1995).  We show that the thermal stability of the hot
one-temperature solution is primarily the result of including
synchrotron emission, which leads to a rapid increase of the cooling
rate with increasing temperature (see Figure \ref{stab}).

An interesting feature of our models is the unusual geometry of the flow.
Since $\dot{M}_{crit}$ decreases with decreasing $R$ in the inner parts 
of the disk (see Figure \ref{mcrit}), for a wide range of $\dot{M}$ the disk 
has a very unusual structure.  In the inner regions, where $\dot{M} > 
\dot{M}_{crit}(R)$, the only possible solution for the accreting gas is 
the Shakura \& Sunyaev thin disk.  Note that since we are restricted to 
relatively low mass accretion rates, the thin disk is viscously stable 
even at $R \sim 3 R_{Schw}$.  Further out, the hot solution exists, but 
it is cooling-dominated and therefore has a disk-like 
geometry.  Finally, in the outer parts of the disk the flow is fully 
advection-dominated and becomes nearly quasi-spherical.  This interesting
result, however, presents a problem for our calculations.  The
inner radiative zone of the disk is cooler than the outer
advection-dominated zone, so the relatively soft photons
emitted on the inside will Compton cool the hot outer gas in a way
that is incompatible with the local treatment of the radiative
processes which we have used in this paper.  The effects of the non-local 
radiation field need to be incorporated self-consistently into future 
calculations.

Since the gas temperatures in our solutions are on the order of $T
\sim 10^{10}$ K, pair production and annihilation processes must be
taken into account.  In our calculations we have included
particle-particle pair production rates (White \& Lightman 1989) which
give values for the pair fraction $z$ less than a few percent, but we
neglected photon-photon and photon-particle processes.  We believe
that this treatment is justified for low values of $\dot{m}$ (low
optical depth), since bremsstrahlung radiation and Comptonization
become unimportant and the cooling is dominated by soft synchrotron
photons which are incapable of producing pairs.  However, for $\dot{M}
\gsim 0.1 \dot{M}_{crit}$, Comptonization of synchrotron radiation
begins to dominate and photon processes may become important.  To
place a limit on these processes we computed the equilibrium pair
density taking into account photon-photon and photon-particle pair
production in the Wien peak of the Comptonized spectrum, using the
rates derived by Svensson (1984).  The resulting values of $z$ are
significantly higher than the values computed taking into account only
particle-particle pair production, as $\dot{M}$ approaches
$\dot{M}_{crit}$.  However, even at $\dot{M}=\dot{M}_{crit}$, the
maximum value of $z$ that we find with the inclusion of photon
interactions is only $\sim{\rm few} \%$.  Kusunose (1996) and Abramowicz 
(1996) confirm that pair processes are unimportant in the hot optically 
thin flows which they have studied.  Thus, we believe that our treatment
of pair processes is reliable in the entire parameter space we have
considered.

\bigskip\bigskip
This work was supported in part by NSF grant AST 9423209 to the Center
for Astrophysics.  RN thanks the Institute for Theoretical Physics
(NSF grant PHY 9407194) for hospitality.  IY acknowledges financial support 
from SUAM Foundation.

\vfill\eject
\begin{appendix}
\section{Comptonization}

Consider a point at an optical depth $\tau_{es}$ inside the medium.
The mean number of scatterings of a soft photon escaping from this
point is (Rybicki \& Lightman 1979)
\begin{equation}
\label{A.1}
s \approx \tau_{es} + \tau_{es}^2.
\end{equation}
Let the average photon energy change per scattering be $A$, i.e. the
initial and final photon energies, $E_{in}$ and $E_{fin}$, are related
on average by $E_{fin} = A E_{in}$.  For a thermal distribution of
electrons with temperature $\theta = k T/(m_e c^2)$, we have (Rybicki
\& Lightman 1979)
\begin{equation}
\label{A.2}
A \approx 1 + 4 \theta + 16 \theta^2.
\end{equation}
The Compton y-parameter is simply $y = s\ (A - 1) = s\ (4 \theta + 16
\theta^2)$.

After $j$ scatterings, the energy gain of the photon is $A^j$ so long
as $E_{fin}$ is not saturated to the Wien regime ($E_{fin} \sim 3 k
T$).  The maximum $j$ corresponding to saturation is thus
\begin{equation}
\label{A.4}
j_m = \ln{\eta_{max}}/\ln{A},\ \ \ \mbox{where}\ \ 
\eta_{max} = \eta_{Wien} = \frac{3 k T}{E_{in}}.
\end{equation}
The energy enhancement factor is then simply the average energy gain
of the photon:
\begin{eqnarray}
\label{A.5}
\eta &=& \sum_{j=0}^{j_m}{A^j P_j} + A^{j_m} \sum_{j = j_m+1}^{\infty}{P_j},
\end{eqnarray}
where $P_j$ is the probability that a photon will suffer exactly $j$ 
scatterings.

To compute $\eta$, we need an expression for $P_j$.  The probability that
a photon travels a distance characterized by an optical depth $\tau_{es}$
without scattering is $e^{-\tau_{es}}$.  To compute $P_j$, the assumption 
is usually made that in a medium with $\tau_{es} \lsim 1$, successive 
scattering events can be treated independently (e.g. Dermer, Liang, \& 
Canfield 1991), so that $P_j$ can be written as $e^{-\tau_{es}} 
(1-e^{-\tau_{es}})^j$.  However, for spherical
geometry the scattering probability decreases after successive scatterings,
because the mean square distance of a photon from the center increases with
each scattering.  Allowing for this accurately requires Monte Carlo 
calculations.  One limit however, is straightforward, namely the case when 
the photon continues to travel radially outward after each scattering.  In 
this case, the probability $P_j$ is given by the Poisson formula
\begin{equation}
\label{A.3}
P_j = \frac{e^{-s} s^j}{j!}.
\end{equation}

With the above expression for $P_j$ we can write 
$\eta$ as
\begin{eqnarray}
\label{A.55}
\eta &=& = e^{-s} \left[\sum_{j=0}^{j_m}{\frac{(A s)^j}{j!}} +  A^{j_m} 
\sum_{j = j_m+1}^{\infty}{\frac{s^j}{j!}}\right] = \nonumber \\ 
&=& e^{-s} \left[e^{As} - \sum_{j=j_m+1}^{\infty}{\frac{(A s)^j}{j!}} +  
A^{j_m} \sum_{j = j_m+1}^{\infty}{\frac{s^j}{j!}}\right].
\end{eqnarray}
To evaluate the sums in equation (\ref{A.5}) we define the following
function:
\begin{equation}
\label{A.6}
f(x) = \sum_{j=j_m+1}^{\infty}{\frac{x^{j}}{j!}}.
\end{equation}
Using integration by parts and the fact that $f(0) = 0$, it is easy to
show that
\begin{equation}
\label{A.7}
f(x) e^{-x} = \frac{1}{j_m!} \int_0^x{e^{-y} y^{j_m} d y} 
\end{equation}
and therefore,
\begin{equation}
\label{A.8}
f(x) = \sum_{j=j_m+1}^{\infty}{\frac{x^{j}}{j!}} = 
\frac{e^{x}}{j_m!} \int_0^x {e^{-y} y^{j_m} d y} = e^{x} P (j_m+1, x),
\end{equation}
where $P (j_m+1, x)$ is the incomplete gamma function.  Substituting
this result into equation (\ref{A.55}) yields the final result:
\begin{eqnarray}
\label{A.9}
\eta & = & e^{-s} \left[\left(e^{A s} - e^{A s} P (j_m+1, A s)\right) + 
A^{j_m} e^s P(j_m+1, s)\right] = \nonumber \\
& = & e^{(A-1) s} \left[1 - P (j_m+1, A s)\right]
+ \eta_{max} P(j_m+1, s).
\end{eqnarray}

\end{appendix}

\vfill\eject
\noindent
{\large \bf References}
\\
\noindent
\def\refpar{\hangindent=3em\hangafter=1}
\def\reference{\refpar\noindent}
\def\apj{ApJ}
\def\apjs{ApJS}
\def\mnras{MNRAS}
\def\aa{A\&A}
\def\aas{A\&AS}
\def\aj{AJ}
\def\nat{Nature}

\reference Abramowicz, M. A. 1996, in Basic Physics of Accretion Disks, ed.
S. Kato, S. Inagaki, J.~Fukue, \& S.~Mineshige (Gordon \& Breach), in press 

\reference Abramowicz, M. A., Chen, X., Kato, S., Lasota, J. P., \& Regev, O. 
1995, \apj, 438, L37

\reference Abramowicz, M. A., Czerny, B., Lasota, J. P., \& Szuszkiewicz, E. 
1988, \apj, 332, 646

\reference Begelman, M. C. \& Chiueh, T. 1988,\apj, 332, 872

\reference Chen, X. 1995, \mnras, 275, 641

\reference Chen, X., Abramowicz, M. A., Lasota, J. P., Narayan, R., Yi, I. 
1995, \apj, 443, L61

\reference Dermer, C. D., Liang, E. P., \& Canfield, E. 1991, \apj, 369, 410

\reference Frank, J., King, A., \& Raine, D. 1992, Accretion Power in 
Astrophysics (Cambridge, UK: Cambridge University press)

\reference Gilfanov, M., et al. 1995, in The Lives of the Neutron Stars, 
NATO ASI Series, eds. M. A. Alpar, U. Kiziloglu \& J. van Paradijs (Kluwer), 
vol. 450, 331

\reference Grebenev, S., et al. 1993, \aas, 97, 281

\reference Harmon, B. A., et al. 1994, The Second Compton Symposium, eds. 
C.~E.~Fichtel, N.~Gehrels \& J.~P.~Norris (New York: AIP), 210

\reference Johnson,W. N., et al. 1994, The Second Compton Symposium, eds. 
C.~E.~Fichtel, N.~Gehrels \& J.~P.~Norris (New York: AIP), 515

\reference Kato, S., Abramowicz, M. A., \& Chen, X. 1995, {\it PASJ}, in press

\reference Kinzer, R. L., et al. 1994, The Second Compton Symposium, eds. 
C.~E.~Fichtel, N.~Gehrels \& J.~P.~Norris (New York: AIP), 531

\reference Kusunose, M. 1996, in Basic Physics of Accretion Disks, eds.
S. Kato, S. Inagaki, J.~Fukue, \& S.~Mineshige (Gordon \& Breach), in press 

\reference Kusunose, M. \& Takahara, F. 1985, Prog. Theor. Phys., 73, 41

\reference Kusunose, M. \& Takahara, F. 1989, PASJ, 41, 263

\reference Lasota, J. P., Abramowicz, M. A., Chen, X., Krolik, J., Narayan, R.,
\& Yi, I. 1996, \apj, 462, 000

\reference Lightman, A. P. \& Eardley, D. M. 1974, \apj, 187, L1

\reference Lynden-Bell, D. \& Pringle, J. E. 1974, \mnras, 168, 603

\reference Luo, C. \& Liang, E. P. 1994, \mnras, 266, 386

\reference Mahadevan, R,  Narayan, R, Yi, I. 1996, \apj, in press

\reference Maisack, M., et al. 1993, \apj, 407, L61

\reference Melia, F. 1994, \apj, 426, 577

\reference Melia, F. \& Misra, R. 1993, \apj, 411, 797

\reference Narayan, R. 1996, \apj, in press

\reference Narayan, R., McClintock, J. E., \& Yi, I. 1996, \apj, in press
(Feb. 1, 1996)

\reference Narayan, R. \& Popham, R. 1993, \nat, 362, 820

\reference Narayan, R. \& Yi, I. 1994, \apj, 428, L13

\reference Narayan, R. \& Yi, I. 1995a, \apj, 444, 231

\reference Narayan, R. \& Yi, I. 1995b, \apj, 452, 277

\reference Narayan, R., Yi, I., \& Mahadevan, R. 1995, \nat, 374, 623

\reference Narayan, R., Yi, I., \& Mahadevan, R. 1996, in Proc. Third Compton
Symposium, \aas\ Special Issue, in press

\reference Novikov, I. D. \& Thorne, K. S. 1973, in Blackholes ed. C. DeWitt
\& B. DeWitt (New York: Gordon and Breach), 343 

\reference Pacholczyk, A. G. 1970, Radio Astrophysics (San Francisco: Freeman)

\reference Phinney, E. S. 1981, in Plasma Astrophysics, eds. T. D. Guyenne \& 
G. Levy (ESA SP-161), 337

\reference Piran, T. 1978, \apj, 221, 652

\reference Pringle, J. E., Rees, M. J., \& Pacholczyk, A. G. 1973, \aa, 29, 
179 

\reference Rees, M. J., Begelman, M. C., Blandford, R. D., Phinney, E. S. 1982,
\nat, 295, 17

\reference Rybicki, G. B. \& Lightman, A. P. 1979, Radiative Processes in 
Astrophysics (New York: John Wiley \& Sons)

\reference Shakura, N. I. \& Sunyaev, R. A. 1973, \aa, 24, 337 

\reference Shapiro, S. L., Lightman, A. P., \& Eardley, D. M. 1976, \apj, 204,
187 

\reference Svensson, R. 1982, \apj, 258, 335

\reference Svensson, R. 1984, \mnras, 209, 175

\reference Tanaka, Y. 1989, in Proc. 23rd ESLAB Symp. on Two Topics in X-Ray 
Astronomy, eds. J. Hunt \& B. Battrick (ESA SP-296), 3

\reference Wandel, A. \& Liang, E. P. 1991, \apj, 380, 84

\reference White, T. R. \&  Lightman, A. P. 1989, \apj, 340, 1024

\newpage

\newpage

\newpage

\newpage

\newpage

\newpage

\newpage

\end{document}